\documentstyle[aps,prl,multicol,floats,epsf]{revtex}

\begin{document}
\draft
\wideabs{

\title{Metal-to-Insulator Crossover in the Low-Temperature 
Normal State of Bi$_{2}$Sr$_{2-x}$La$_x$CuO$_{6+\delta}$}

\author{S. Ono,$^{1}$ Yoichi Ando,$^{1,2,}$\cite{YA}
T. Murayama,$^{1,2,}$\cite{Mura} F. F. Balakirev,$^{3}$
J. B. Betts,$^{3}$ and G. S. Boebinger$^{3}$}
\address{$^1$Central Research Institute of Electric Power
Industry, Komae, Tokyo 201-8511, Japan}
\address{$^{\rm 2}$ Department of Physics, Science University of Tokyo,
Shinjuku-ku, Tokyo 162-8601, Japan}
\address{$^{\rm 3}$NHMFL, Los Alamos National Laboratory, Los Alamos,
NM 87545}

\date{\today}
\maketitle

\begin{abstract}
We measure the normal-state in-plane resistivity of
Bi$_{2}$Sr$_{2-x}$La$_x$CuO$_{6+\delta}$ single crystals
at low temperatures by suppressing superconductivity
with 60-T pulsed magnetic fields.
With decreasing hole doping, we observe a crossover from a
metallic to insulating behavior in the low-temperature normal state.
This crossover is estimated to occur near 1/8 doping, well inside the 
{\it underdoped} regime, and not at optimum doping as reported for 
other cuprates.
The insulating regime is marked by a logarithmic temperature dependence of
the resistivity over two decades of temperature, suggesting that a peculiar
charge localization is common to the cuprates.  
\end{abstract}

%74.25.Dw Superconductivity phase diagrams
%74.25.Fy Transport properties (electric and thermal conductivity, etc.)
%74.72.Hs Bi-based cuprates
}
\narrowtext

The normal-state of the high-$T_c$ cuprate superconductors exhibits
unusual properties thought to evidence non-Fermi-liquid behavior
\cite{NFL}.
A sufficiently-intense magnetic field can suppress the superconducting
phase and reveal the low-temperature limiting behavior in this
strongly-correlated electron system \cite{Ando,GSB}.
For example, a metal-to-insulator (MI) crossover
takes place at optimum doping in the normal state of the hole-doped
La$_{2-y}$Sr$_{y}$CuO$_{4}$ (LSCO) system \cite{GSB},
implying that superconductivity in the underdoped regime may be
realized in an otherwise-insulating system.

Recently, it was reported that a MI crossover also takes place at
optimum doping in the electron-doped Pr$_{2-x}$Ce$_{x}$CuO$_{4+\delta}$
(PCCO) system \cite{Fournier}.  If a MI crossover near optimum doping
is a universal feature of all the cuprates,
one plausible conclusion is that
there is a quantum critical point at optimum doping, which might be the
source of the non-Fermi-liquid
behavior in the cuprates \cite{QCP}.
To date, however, only two systems have been systematically studied.

Surprisingly, the insulating behavior in both LSCO and PCCO
is characterized by the in-plane resistivity $\rho_{ab}(T)$
increasing as $\log (1/T)$ \cite{Ando,Fournier,JLTP}.
This apparently diverging resistivity is sufficiently weak that it is
experimentally difficult to precisely establish its functional form.
Measurements on LSCO currently offer the
best experimental case for a strict $\log (1/T)$ behavior with no
evidence
of saturation at low temperatures \cite{Ando,JLTP}.
The data from PCCO do not span such a large dynamic range and tend
toward saturation below $\sim$3 K \cite{Fournier}.
Because the $\log (1/T)$ divergence is not attributable to any known
localization mechanism \cite{Ando,JLTP,AndoHall},
it attracts substantial theoretical interest.
Questions remain regarding its universality among the cuprates.

Bi$_{2}$Sr$_{2-x}$La$_x$CuO$_{6+\delta}$ (BSLCO, or La-doped Bi-2201)
is well-suited for a systematic study of the low-temperature
normal state because the carrier concentration can be tuned in
both the overdoped and underdoped regimes
\cite{Maeda,Murayama}. Also, since the maximum $T_c$ of BSLCO is
relatively low, superconductivity can be completely suppressed with
experimentally-available 60-T pulsed magnetic fields.
By suppressing superconductivity in a series of BSLCO single crystals,
we measure $\rho_{ab}$ in the normal state down to
$\sim$0.5 K.  We find that in BSLCO the MI crossover, or the onset of
localization, takes place not at optimum doping but well inside the
underdoped regime.  Hall measurements support the conclusion that the
MI crossover occurs in BSLCO at approximately 1/8 hole concentration.

The single crystals of BSLCO are
grown
using a floating-zone technique over a wide range of La concentration
$x$ = 0.23, 0.39, 0.66, 0.73, 0.76 and 0.84.
Note that a larger $x$ yields a
more underdoped sample and $x$=0.39 is optimal-doping.
The actual La concentration is measured by
inductively-coupled plasma analysis and La homogeneity is
confirmed by electron-probe microanalysis.
For transport measurements, samples of
$\sim$ 2 $\times$ 1 $\times$ 0.05 mm$^{3}$ are cut from large crystals.
The thickness is accurately determined by
weighing the samples with 0.1-$\mu$g resolution, which reduces the
uncertainty in the absolute magnitude of $\rho_{ab}$ to $\pm$5\%
which results from the finite contact size.
All the samples are annealed in flowing oxygen at 650$^\circ$C for 48
hours to sharpen the superconducting transition, whose width (from
dc magnetic susceptibility) is typically 2 K after annealing.
The oxygen annealing also helps raise the optimum $T_c$;
the optimally-doped sample reported here shows zero-resistance at 38 K,
the highest value ever reported for BSLCO
and almost equal to the optimum $T_c$ of LSCO.

The magnetic field dependence of $\rho_{ab}(H)$ is measured at fixed
temperatures with the same $\sim$100 kHz four-probe technique described
previously \cite{Ando,GSB}.  The pulsed magnetic field is applied
parallel to the $c$ axis to best suppress superconductivity.
Two different pulsed magnets were utilized in these experiments:
a 60-T magnet with a 15 msec pulse duration and a ``long-pulse"
40-T magnet with a 500 msec pulse duration.  Eddy-current heating,
proportional to $(dH/dt)^{2}$, does not adversely affect the data,
as confirmed by comparing data from the two different magnets and from
pulses with different peak fields.

\begin{figure}[t!]
\epsfxsize=0.85\columnwidth
\centerline{\epsffile{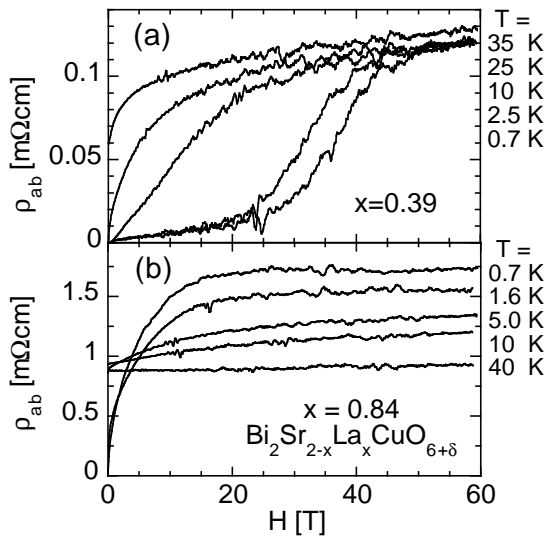}}
\vspace{0.2cm}
\caption{
Suppression of superconductivity by intense magnetic fields in BSLCO
crystals with (a) $x$=0.39 and (b) $x$=0.84 at selected temperatures.}
\label{fig1}
\end{figure}

Figure 1 shows representative traces of $\rho_{ab}(H)$ at selected
temperatures for $x$ = 0.39, the optimally-doped sample, and $x$ = 0.84,
the most underdoped sample studied. Note that 60-T
is sufficient to suppress superconductivity in BSLCO,
even at optimum doping. Because the magnetoresistance in the
normal state is negligible, we consider that the 60-T data represents
the normal-state behavior at temperatures below $T_c$.

\begin{figure}[t!]
\epsfxsize=0.9\columnwidth
\centerline{\epsffile{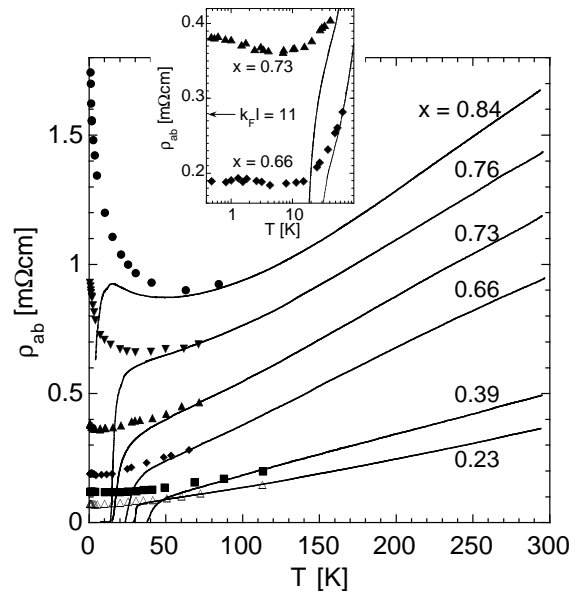}}
\vspace{0.2cm}
\caption{
Temperature dependence of $\rho_{ab}$ for the BSLCO crystals in 0 T
(solid lines) and in 60 T (symbols) for different La concentration $x$.
The inset offers a clearer view of the low-temperature behaviors of the
$x$=0.73 and 0.66 samples. }
\label{fig2}
\end{figure}

Figure 2 shows the $T$ dependence of $\rho_{ab}$ in 0 and 60 T for
the six $x$ values. In zero field (solid lines), all samples
are metallic ($d\rho_{ab}/dT \ge$ 0) above $T_c$ except for the
$x$=0.84 sample, in which the
carrier concentration is sufficiently low that the superconducting
phase has
nearly disappeared.  However, the 60-T data (symbols) reveal that the
two most heavily underdoped samples, $x$=0.84 and 0.76, both show a
strong increase in $\rho_{ab}$ at low temperatures.
As is most clear in the inset to Fig. 2, the $x$=0.73 sample shows a
very weak upturn below 6K, which we interpret
as the onset of localizing behavior in BSLCO. In contrast,
$\rho_{ab}(T)$
for the $x$=0.66 sample is essentially constant below 10 K.
The weak upturn in the $x$=0.73 sample makes it difficult to
precisely determine the La concentration at the boundary for the
onset of localization in the low-temperature normal state of BSLCO;
however, clearly metallic behavior is found at $x$=0.66 in an
underdoped sample with $T_c$ = 23 K
and, thus, the MI crossover in BSLCO appears to lie in the
underdoped regime, not at optimum doping as in LSCO and PCCO.

Figure 3 contains selected BSLCO data from Fig. 2 replotted (filled
circles) versus the logarithm of the temperature. Note that the
low-temperature data for $x$=0.84 increases as $\log (1/T)$ with
decreasing
temperature. To determine whether this pronounced $\log (1/T)$ behavior
persists to our lowest achievable temperature, we re-measured
this sample at $H$=30 T at 1.5 K and 0.3 K (open circles) using the 40 T
long-pulse magnet. In the $x$ = 0.84 sample, 30 T is sufficient to
suppress superconductivity (Fig. 1b).
These 30-T data are consistent with
those from the 60-T pulses and the $\log (1/T)$ behavior is found to
extend over two decades of temperature
from 30 K to 0.3 K without any sign of saturation at low-temperatures.
This constitutes evidence for a strictly $\log (1/T)$ divergence that is
even stronger than has been reported in LSCO \cite{Ando}. As in LSCO,
this $\log (1/T)$ behavior occurs at resistivities on the metallic
side of the Mott limit ($k_F l$=1 corresponds to
$\rho_{ab}$$\simeq$3.1 m${\rm \Omega}$cm in BSLCO).

\begin{figure}[t!]
\epsfxsize=0.9\columnwidth
\centerline{\epsffile{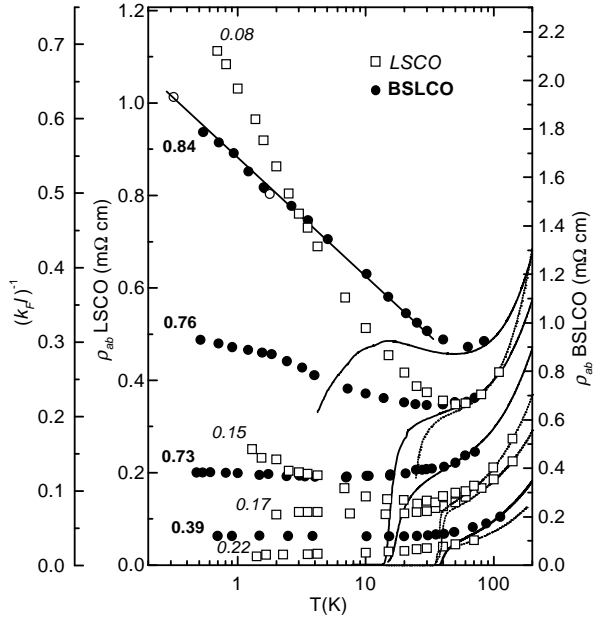}}
\vspace{0.2cm}
\caption{
Logarithmic plot of $\rho_{ab} (T)$ for BSLCO crystals in 0 T (solid
lines)
and in 60 T (filled circles), labelled by
La concentration, $x$. The straight line emphasizes
the $\log(1/T)$ behavior in the $x$=0.84 sample and open circles are
30-T data from the long-pulse magnet. LSCO data in 0 T (dashed lines)
and in 60 T (open squares), labelled by Sr concentration, $y$, are
from Ref. \protect\cite{GSB}.  All data are vertically scaled
to directly compare the resistivity per CuO$_2$ layer in units
of $(k_{F}  l)^{-1}$.}
\label{fig3}
\end{figure}

Figure 3 also includes $\rho_{ab}(T)$ data from LSCO (open squares),
reproduced from Ref. \cite{GSB}.  The vertical axes used for
plotting the
LSCO and BSLCO data have been scaled to directly compare the
resistivity per CuO$_2$ layer, $\rho_{ab} / c_0$, where $c_0$ is
the $c$-axis lattice spacing.  This normalized resistivity is given
in units of $(k_F l )^{-1}$ on the left-most axis:
$\rho_{ab} / c_0 = h / ( e^2 k_{F}  l)$, where $k_F$ is the Fermi wave
vector, and $l$ is the mean free path \cite{Ando2201}.
Several points are worth noting:
(a) Optimally-doped BSLCO ($x$=0.39) exhibits a lower normalized
resistivity than optimally-doped LSCO.  This might account for the
strongly insulating
behavior beginning further from optimal-doping in BSLCO than in LSCO.
(b) The LSCO and BSLCO data are consistent with the MI crossover
occurring at a common value of normalized $\rho_{ab}$ in the
low-temperature limit, corresponding to $k_F l \simeq$ 12, if the
$x$=0.73 data are interpreted as insulating behavior.
(c) The MI crossover occurs much more
abruptly in LSCO (between $y$ = 0.15 and 0.17) than in BSLCO.
In BSLCO, the onset of localizing behavior occurs relatively
gradually as the
normalized $\rho_{ab}(T)$ becomes increasingly larger with increasing
La concentration.
(d)  When comparing BSLCO and LSCO samples exhibiting similar
normalized resistivities, the insulating behavior is stronger in
LSCO than in BSLCO.
This suggests that the normalized resistivity and estimated magnitude of
$k_F l$ are not the sole factor determining the strength of the $\log
(1/T)$
behavior in underdoped cuprates.

Developing a phase diagram for BSLCO is difficult because the
hole concentration per Cu, $p$, cannot be chemically determined.
Even if the La concentration, $x$, and excess oxygen,
$\delta$, are accurately known, the hole concentration $p$ cannot be
calculated because Bi-ion does not have a fixed valency \cite{Idemoto}.
We therefore estimate $p$ for our BSLCO crystals based on Hall
coefficient $R_H$.
Figure 4 compares $R_H$ from three different single-layer cuprates
by normalizing with the unit cell volume $V$. Both the magnitude and
the $T$-dependence of
$R_H/V$ of our optimally-doped BSLCO agree reasonably well with the
optimally-doped ($y$=0.15) LSCO data \cite{Kimura} and the
optimally-doped Tl$_2$Ba$_2$CuO$_{6+\delta}$ data \cite{Kubo}.
From this, we conclude that our optimally-doped $x$=0.39 BSLCO sample
has a hole concentration $p \simeq$ 0.15.
In LSCO, the $p$ value is equal to the Sr concentration, $y$,
which is used as a guide to estimate $p$ for each of our BSLCO
samples on the basis of the comparison of the $R_H/V$ data.
The approximate relationship between La concentration, $x$,
and hole concentration, $p$, is given in Fig. 5.

\begin{figure}[t!]
\epsfxsize=0.85\columnwidth
\centerline{\epsffile{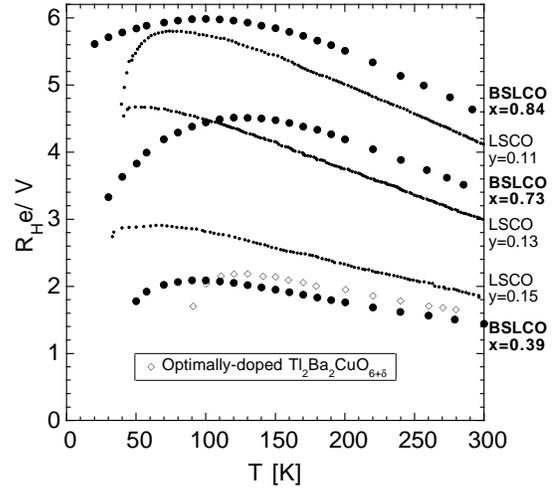}}
\vspace{0.2cm}
\caption{
Normalized Hall coefficient for several of our
BSLCO samples, where $e$ is the electronic charge and $V$ is the unit
cell volume.  LSCO data are from Ref.
\protect\cite{Kimura} and optimally-doped Tl$_2$Ba$_2$CuO$_{6+\delta}$
data are from Ref. \protect\cite{Kubo}. }
\label{fig4}
\end{figure}

Figure 5 contains the phase diagram resulting from our BSLCO transport
data
of Figs. 2 - 4. The zero-resistance superconducting transition
temperatures $
T_{c}$ for our BSLCO samples is plotted (filled circles) versus
La concentration $x$.  The MI boundary, or the onset of the localizing
behavior, in the normal state is schematically denoted by the open
diamonds,
the temperatures $T_{min}$ where the $\rho_{ab}(T)$ shows a minimum in
the
60-T data of Fig. 2.  The top axis contains the approximate hole
concentration
$p$ deduced from the Hall data of Fig. 4. It is clear from the phase
diagram
in Fig. 5 that the MI crossover at zero temperature lies well inside the
underdoped regime for BSLCO, at a La concentration of $x$ $\simeq$ 0.7.
The hole concentration at the MI crossover is estimated to be
$p$ $\simeq$ 1/8.

Note that Fig. 5 suggests that the superconducting phase in
BSLCO extends only down to $p \sim $ 0.10, instead of $p \sim$ 0.05
as with other hole-doped cuprates. This implies that the $x$=0.84
sample exhibits such a low $T_c$ not because $p \sim$ 0.05 but because
$T_c$ is reduced by disorder. This conclusion is supported by
room-temperature thermopower measurements which are useful
to independently estimate $p$ in the cuprates \cite{Tallon}.
We measure the thermopower of the BSLCO $x$=0.84 sample to be
38 $\mu$V/K at room temperature, a value much smaller than the
$\sim$80 $\mu$V/K that would be expected at $p \sim$ 0.05
\cite{Tallon}.

Although the role of disorder is difficult
to characterize, we note that disorder enhances insulating 
behavior in LSCO \cite{JLTP} and argue that increased disorder 
will likely expand the insulating regime.  
In other words, even though BSLCO crystals are disordered, 
less-disordered crystals are likely to exhibit a smaller 
insulating regime that recedes further into the underdoped regime. 
Hence, we believe the intrinsic behavior in BSLCO is that 
optimum doping occurs at $p \simeq$ 0.15 and the insulating regime 
does not extend to optimum doping \cite{note}.

\begin{figure}[t!]
\epsfxsize=0.8\columnwidth
\centerline{\epsffile{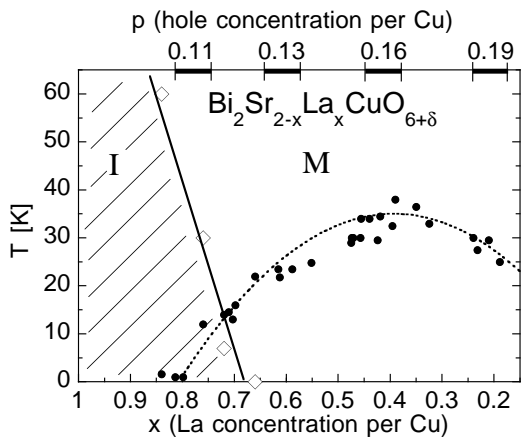}}
\vspace{0.2cm}
\caption{
Schematic phase diagram of BSLCO showing the superconducting
$T_{c}$ (solid circles) and $T_{min}$ (open diamonds), the temperature
at the minimum in $\rho_{ab}$.
The solid line is boundary separating the insulating (I)
and metallic (M) regimes. The top axis gives the approximate
hole concentration corresponding to the measured La concentration.}
\label{fig5}
\end{figure}

In LSCO, localizing behavior extends throughout the entire 
underdoped regime up to optimum doping \cite{GSB}.
Neutron scattering data on LSCO \cite{Yamada} suggests 
that charge is confined to stripes over the same range 
of carrier concentration: a plot of $T_c$ versus 
incommensurability $\delta$ (Fig. 10 of Ref. \cite{Yamada} ) 
finds a linear relation between $T_c$ and $\delta$ that 
persists to $p$=0.15. 
Therefore, it is tempting to speculate that the 
$\rho_{ab} \sim \log (1/T)$ insulating behavior in LSCO 
might be associated with low dimensionality imposed by 
charge confinement onto stripes. 
If so, our transport data on BSLCO suggest that neutron 
scattering experiments will find the incommensurability, 
or the inverse separation between the stripes, 
to saturate for $p > $ 0.12 in BSLCO.  
Additional transport studies on other hole-doped cuprates 
might further help establish general correlations between 
hole doping, stripe formation, and the unusual $\log (1/T)$ 
divergence of $\rho_{ab}$.

We note that the charge-stripe instability is often thought to be 
tied to the proximity to an antiferromagnetic ground state.  
A N\'{e}el transition has indeed been reported for 
heavily-La-doped Bi-2201 \cite{Maeda}.  
Also, recent theoretical arguments have linked the 
relatively low value of $T_c$ at optimum doping in Bi-2201 
to a proposed charge-stripe instability \cite{Baskaran}. 
These are additional reasons to seek still-missing direct 
experimental evidence for stripes in BSLCO or Bi-2201, 
which might correlate with our observed insulating phase. 

To summarize, we measure the in-plane resistivity of 
BSLCO crystals down to low temperatures by suppressing 
superconductivity with 60-T pulsed magnetic fields. 
We find that metallic behavior gradually changes to 
insulating behavior with decreasing carrier concentration.
This metal-to-insulator crossover in BSLCO takes place in
the underdoped regime, at hole doping $p \sim$ 1/8,
in striking contrast to the behavior reported for LSCO and PCCO.
Thus, the MI crossover in cuprates is not
universally associated with optimum doping.  
On the other hand, $\rho_{ab}$ in the insulating regime 
displays the clearest $\log(1/T)$ temperature dependence 
yet observed, suggesting a peculiar charge localization that is 
common in all the cuprates.

% Place here the list of the references:
%
\medskip
\vfil


\begin{references}
\bibitem{note}
One might speculate that disorder is distorting the 
BSCLO phase diagram, by preferentially suppressing 
superconductivity in the underdoped regime 
due to increased disorder there.
If so, in cleaner samples, the maximum $T_c$ might occur 
at $p \simeq 0.12$ and the MI crossover and ``optimal doping" 
could coincide.  
On the other hand, we note that our interpretation, 
based upon our BSLCO Hall data, supports one of the most 
striking commonalties of the hole-doped cuprates: 
that optimal doping occurs at $p \simeq 0.15$.
\end{references}
\end{document}